\documentclass[twocolumn,trackchanges]{aastex631}
\usepackage{amsmath}

\begin{document}
\title{Identification of OB associations using the LAMOST-Gaia OB star sample}

\author[0000-0001-5314-2924]{Zhicun Liu}
\altaffiliation{Physics Postdoctoral Research Station at Hebei Normal University}
\affiliation{Department of Physics, Hebei Normal University, Shijiazhuang 050024, People's Republic of China}
\affiliation{Guo Shoujing Institute for Astronomy, Hebei Normal University, Shijiazhuang 050024, People's Republic of China}
\affiliation{Hebei Key Laboratory of Photophysics Research and Application, Shijiazhuang 050024, People's Republic of China}
\affiliation{Shijiazhuang Key Laboratory of Astronomy and Space Science}

\author[0000-0003-2536-3142]{Xiao-Long Wang}
\altaffiliation{Physics Postdoctoral Research Station at Hebei Normal University}
\affiliation{Department of Physics, Hebei Normal University, Shijiazhuang 050024, People's Republic of China}
\affiliation{Guo Shoujing Institute for Astronomy, Hebei Normal University, Shijiazhuang 050024, People's Republic of China}
\affiliation{Shijiazhuang Key Laboratory of Astronomy and Space Science}
\affiliation{Hebei Advanced Thin Films Laboratory, Shijiazhuang 050024, People's Republic of China}

\author[0000-0003-1359-9908]{Wenyuan Cui}
\affiliation{Department of Physics, Hebei Normal University, Shijiazhuang 050024, People's Republic of China}
\affiliation{Guo Shoujing Institute for Astronomy, Hebei Normal University, Shijiazhuang 050024, People's Republic of China}
\affiliation{Shijiazhuang Key Laboratory of Astronomy and Space Science}

\author[0000-0002-0349-7839]{Jianrong Shi}
\affiliation{CAS Key Laboratory of Optical Astronomy, National Astronomical Observatories, Chinese Academy of Sciences, Beijing 100101, People's Republic of China}

\author[0000-0002-1802-6917]{Chao Liu}
\affiliation{Key Laboratory of Space Astronomy and Technology, National Astronomical Observatories, Chinese Academy of Sciences, Beijing 100101, People's Republic of China}
\affiliation{Institute for Frontiers in Astronomy and Astrophysics, Beijing Normal University, Beijing 102206, People's Republic of China}

\author[0000-0002-0642-5689]{Xiang-Xiang Xue}
\affiliation{CAS Key Laboratory of Optical Astronomy, National Astronomical Observatories, Chinese Academy of Sciences, Beijing 100101, People's Republic of China}

\author[0000-0001-8060-1321]{Min Fang}
\affiliation{Purple Mountain Observatory, Chinese Academy of Sciences, 10 Yuanhua Road, Nanjing 210023, China}

\author[0000-0001-8424-1079]{Yang Huang}
\affiliation{CAS Key Laboratory of Optical Astronomy, National Astronomical Observatories, Chinese Academy of Sciences, Beijing 100012, People's Republic of China}

\author[0000-0003-1828-5318]{Guozhen Hu}
\altaffiliation{Physics Postdoctoral Research Station at Hebei Normal University}
\affiliation{Department of Physics, Hebei Normal University, Shijiazhuang 050024, People's Republic of China}
\affiliation{Guo Shoujing Institute for Astronomy, Hebei Normal University, Shijiazhuang 050024, People's Republic of China}
\affiliation{Hebei Key Laboratory of Photophysics Research and Application, Shijiazhuang 050024, People's Republic of China}
\affiliation{Shijiazhuang Key Laboratory of Astronomy and Space Science}
\author[0000-0002-8980-945X]{Gang Zhao}
\affiliation{CAS Key Laboratory of Optical Astronomy, National Astronomical Observatories, Chinese Academy of Sciences, Beijing 100101, People's Republic of China}

\correspondingauthor{Wenyuan Cui}
\email{cuiwenyuan@hebtu.edu.cn, wenyuancui@126.com}
\submitjournal{\apjs}


\begin{abstract}

OB associations, as an intermediate stage between Galactic clusters and field stars, play an important role in understanding the star formation process, early stellar evolution, and Galactic evolution. In this work, we construct a large sample of OB stars with 6D phase space parameters ($l, b, d, V_{\rm los}, pmra ,pmdec$) by combining the distances from \citet{Bailer2021AJ....161..147B}, radial velocities derived from low-resolution spectra of the Large Sky Area Multi-Object Fiber Spectroscopic Telescope (LAMOST), and proper motions from the \textit{Gaia} Data Release 3 (DR3). This sample includes 19,933 OB stars, most of which are located within 6\,kpc of the Sun. Using 6D phase space parameters and friends-of-friends clustering algorithm, we identify 67 OB associations and 112 OB association candidates, among them, 49 OB associations and 107 OB association candidates are newly identified. The Galactic rotation curve derived using 67 OB association members is relatively flat in the range of Galactocentric distances 7$<$$R$$<$13\,kpc. The angular rotation velocity at solar Galactocentric distance of $R_\odot$ =8.34\,kpc is $\Omega_0$ = 29.05$\pm$0.55\,km\,s$^{-1}$\,kpc$^{-1}$. The spatial distribution of the 67 OB associations indicates that they are mainly located at low Galactic latitudes and near spiral arms of the Milky Way. Additionally, we estimate the velocity dispersions and sizes of these 67 OB associations. Our results show that the velocity dispersions decrease as Galactocentric distances increase, while their sizes increase as Galactocentric distances increase.



\end{abstract}



\section{introduction}

OB associations are gravitationally unbound stellar groups including numerous O-type and B-type stars, whose total masses usually range from a few thousand to tens of thousands of solar masses \citep{Ambartsumian1947esa..book.....A,deZeeuw1999AJ....117..354D,Wright2023ASPC..534..129W}. They typically spread tens to hundreds of parsecs and contain smaller groups with distinct kinematics \citep{Magnier1993A&A...278...36M,Garmany1994PASP..106...25G,mel2017MNRAS.472.3887M}. OB associations are closely linked to the star-forming regions and spiral arms in the Milky Way because of the short evolutionary time scale of OB stars \citep{Wright2020NewAR..9001549W}. Consequently, OB associations are excellent tracers for studying star formation regions, the star formation process, and the initial mass function of galaxies.  

There are two primary scenarios that explain the origins of OB associations: (i) the clustered model of star formation: massive stars form mainly in stellar clusters, and the OB associations are the expanded remnant of the dense cluster's parent molecular cloud that was destroyed during a feedback phase \citep{Blaauw1964ARA&A...2..213B,Lada2003ARA&A..41...57L,Alexis2023MNRAS.522.3124Q}. (ii) the hierarchical star formation model: the stars form in groups with different densities and sizes, and they are unbound and disperse from birth \citep{Kruijssen2012MNRAS.426.3008K,Alexis2021MNRAS.508.2370Q}. The key difference between those two suggestions is whether there is an expansion pattern of OB associations. The typical expansion velocity of OB associations is about 5\,km s$^{-1}$ to balance between their initial expansion and the Galactic tidal forces \citep{Ambartsumian1949AZh....26....3A}. Using the data from \textit{Gaia} DR2 (\citet{2018A&A...616A...1G}), \citet{Kounkel2018AJ....156...84K} found significant expansion in the Orion D group by studying its kinematics, and \citet{Melnik2020MNRAS.493.2339M} confirmed that only six of 28 OB associations are expanding by investigating their internal motions. However, \citet{Wright2018MNRAS.476..381W} found that there is no evidence for the expansion of three subgroups in the Scorpius-Centaurus (Sco-Cen) OB association (Sco OB2) with their kinematic study obtained by using the \textit{Gaia} DR1 parallaxes and proper motions. \citet{Ward2018MNRAS.475.5659W} also found that the 18 OB associations identified from the Tycho–Gaia Astrometric Solution catalog do not show evidence of expansion.  
Generally, most stars in star-forming regions are believed to form in clusters \citep{Porras2003AJ....126.1916P,Koenig2008ApJ...688.1142K}. However, based on the observed number of clusters compared to the predictions of a constant cluster formation rate model, less than 10 percent of these clusters remain bound \citep{Lada2003ARA&A..41...57L}. OB associations, which are unbound groups containing large samples of young stars, represent a transitional phase between star clusters and the field star population in galaxies. They are invaluable for investigating the star formation process, the origins of the field star population, and the properties of multiple systems \citep{Kroupa2011sca..conf...17K,Wright2020NewAR..9001549W}.

\begin{figure}[htp!]
    \centering
    \includegraphics[width=1.0\linewidth]{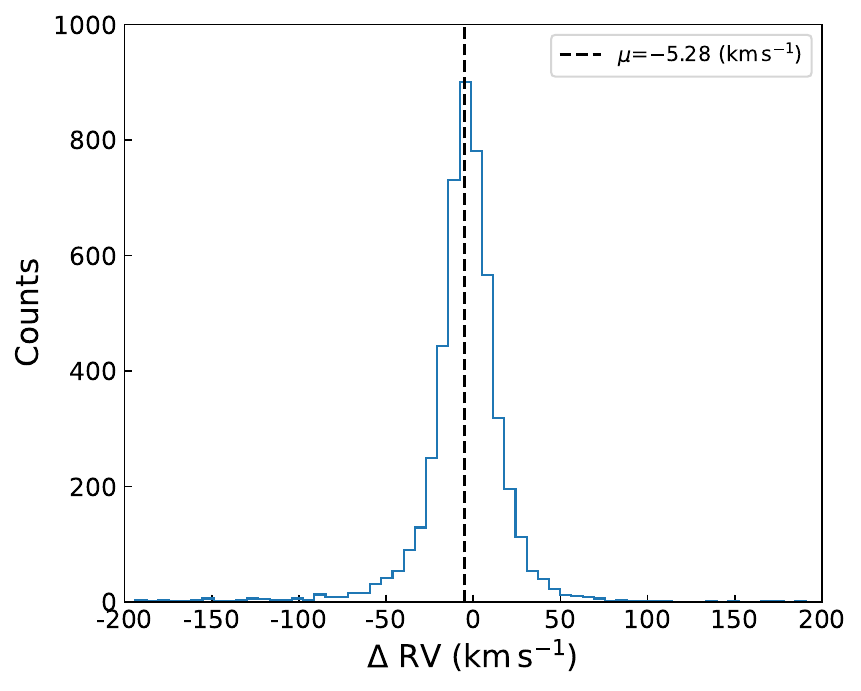}
    \caption{Distribution of the values of $\Delta$\,RV\,=RV$_{\rm This \ work}$-RV$_{\rm \textit{Gaia} \ DR3}$ for 4939 common OB stars. The black dashed line represents the mean value ($\mu$).}
    \label{FigureRV}
\end{figure}

Thanks to \textit{Gaia} data, many OB associations have been identified and revisited. For example, \cite{Ward2020MNRAS.495..663W} identified 109 likely OB associations using a clustering algorithm, based on a sample of 11,844 Galactic OB stars with the \textit{Gaia} DR2 data. In addition, \citet{Alexis2021MNRAS.508.2370Q} identified six new OB associations by reviewing the Cygnus OB association, while \citet{Chemel2022MNRAS.515.4359C} applied the HDBSCAN algorithm \citep{Campello2013HDBSCAN,Campello2015HDBSCAN,McInnes2017JOSS....2..205M} to search for OB associations in \textit{Gaia} EDR3. 

In this paper, we aim to identify OB associations using OB stars from LAMOST DR7 and \textit{Gaia} DR3. In Section~\ref{sec:data}, we describe how to obtain the data sample. The methods used to identify OB association are introduced in Section~\ref{sec:method}. In Section~\ref{sec:RandD}, we present the results and discussion. A summary is provided in Section~\ref{sec:conclusion}.

\section{Data Sample} \label{sec:data}

\begin{figure*}
    \centering
    \includegraphics[width=0.9\textwidth]{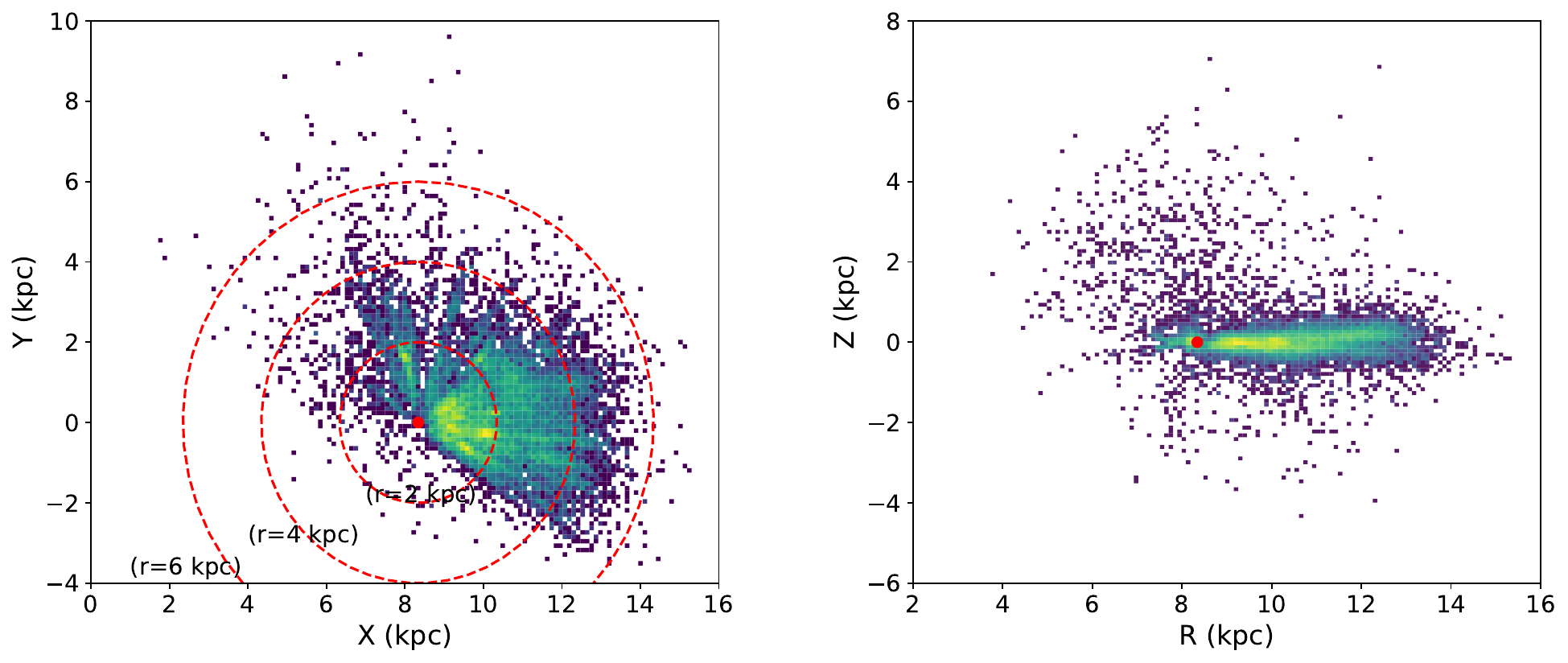}
    \caption{Distribution of 19,933 OB stars in the Galactic X-Y (left panel) and R-Z (right panel) plane. The black dots represent the solar position (X=8.34\,kpc, Y=0\,kpc, Z=0\,kpc) obtained from \citet{Reid2014ApJ...783..130R}. The dashed-red rings in the left panel delineate constant distances from the Sun in steps of 2\,kpc.}
    \label{Fig01XYZ}
\end{figure*}

\begin{figure*}
    \centering
    \includegraphics[width=1.0\textwidth]{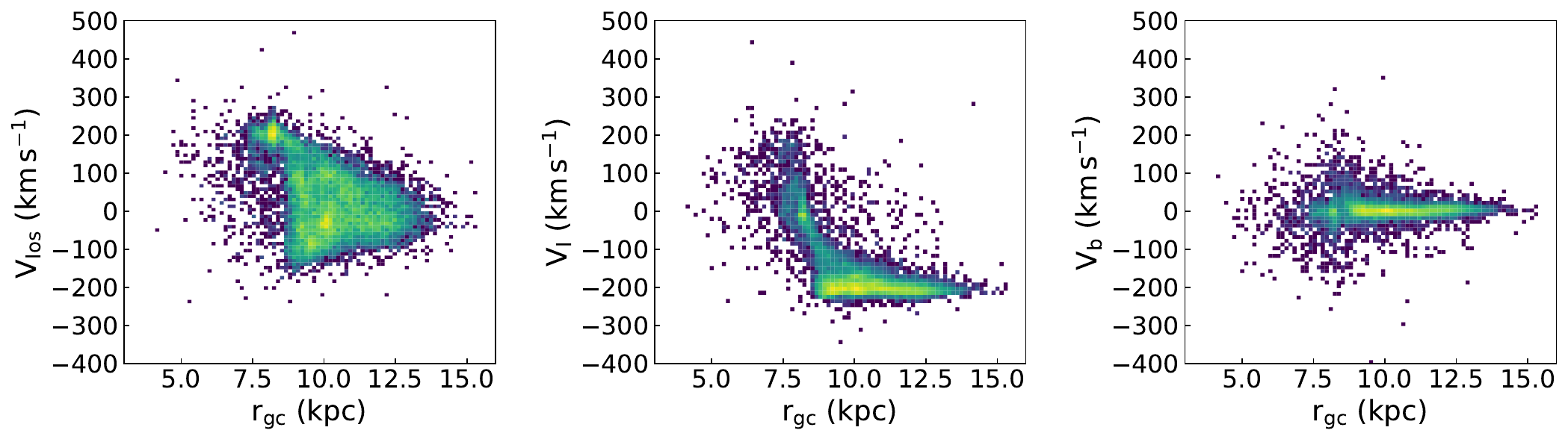}
    \caption{The velocity distributions along with the Galactocentric distance r$_{\rm gc}$ of our 19,933 OB stars. }
    \label{Fig02VlVb}
\end{figure*}

In this work, we use the OB stars selected by \citet{Liu2024ApJS..275...24L} from LAMOST DR7 to identify OB associations. The catalog includes 37,778 spectra of 27,643 OB stars identified in the line index's space using improved OB star selection criteria \citep{Liu2019ApJS..241...32L,Liu2024ApJS..275...24L}. In addition, we include 137 OB stars that are missed due to their low signal-to-noise ratios of the low-resolution spectra. Thus, the final OB star sample consists of 27,780 stars (27,643+137). After cross-matching with \textit{Gaia} DR3 and the distance catalog of \citet{Bailer2021AJ....161..147B} with a matching radius of 3 arcsecs, we obtained the parallaxes, proper motions, RUWE, and the photometric distances for 26,877 OB-type stars. Although \citet{Bailer2021AJ....161..147B} claimed that the distances of distant giants are underestimated, this effect is negligible because our OB star sample primarily includes main-sequence stars. Considering the errors in the \textit{Gaia} astrometric data, we only adopted OB stars that satisfy the following criteria:

\begin{itemize}
\item[(i)]$\frac{\rm Parallax\_error}{\rm Parallax}<20\%$ and RUWE$<$1.4.
\item[(ii)] Removing the stars without proper motions.  
\end{itemize}

After applying the above criteria, a total of 19,933 OB stars have been selected.

The radial velocity is calculated using the cross-correlation-based $laspec$ algorithm from \citep{Zhang2021ApJS..256...14Z}, based on BOSZ synthetic spectra library \citep{Bohlin2017AJ....153..234B,Meszaros2012AJ....144..120M}. Considering the effective temperatures, surface gravities, metallicity, and projected rotational velocities ($v$\,sin\,$i$) of OB stars, we download the model spectra with a resolution of 5000, solar metallicity, and an effective temperature greater than 8000\,K from the website \footnote{https://archive.stsci.edu/prepds/bosz/}. We then degrade the spectra to the LAMOST resolution (R$\sim$1800) with different $v$\,sin\,$i$ \footnote{Given the LAMOST spectra resolution, we chose $v$\,sin\,$i$ values of 10, 50,100,150,200,250,300,350,400,450,500\,km s$^{-1}$}. It is applied to the spectra of OB stars with wavelengths between 3850 and 4950\,\AA~ from LAMOST DR7. Moreover, the radial velocities of Be-type stars with H$_\beta$ emission lines are measured using the H$_\beta$-masked spectrum. Previous results indicate that stellar radial velocities derived from LAMOST low-resolution spectra are underestimated \citep{Huang2018AJ....156...90H,Wang2021MNRAS.504..199W}. Therefore. We cross-match the 19,933 OB stars with the \textit{Gaia} DR3 data and find that there are 4939 common stars with radial velocities. Figure \ref{FigureRV} shows the distribution of $\Delta$ RV=RV$_{\rm This \ work}$-RV$_{\rm \textit{Gaia} \ DR3}$ for 4939 common OB stars. We find that the radial velocities from this work are smaller than those from \textit{Gaia} DR3 by 5.28 km$\cdot$s$^{-1}$, and add 5.28 km$\cdot$s$^{-1}$ to the radial velocities of our sample as a correction. 

Here, we use the right-handed Cartesian coordinate centered at the Galactic center. The X-axis is positive toward the Galactic Center from the Sun, the Y-axis is along the rotation of the Galactic disk, and the Z-axis points toward the North Galactic Pole. The solar position is located at ($-8.34, 0, 0$) kpc \citep{Reid2014ApJ...783..130R}. Using the radial velocity, distance, and proper motions of OB stars, we calculate their velocities that are converted to the Galactic standard of rest (GSR) frame by adopting the solar motion of ($+11.1$, $+12.24$, $+7.25$) km s$^{-1}$ with respect to the local standard of rest (LSR) frame \citep{schonrich2010MNRAS.403.1829S}, the LSR rotational velocity of 220 km$\cdot$s$^{-1}$ \citep{Kerr1986MNRAS.221.1023K}. Finally, a sample of 19,933 OB stars with 3D positions and 3D velocities is constructed. Figure~\ref{Fig01XYZ} shows the distribution of OB stars in the Galactic X-Y (left panel) and X-Z (right panel) planes. The velocity distributions of OB stars with different Galactocentric distances are shown in Figure~\ref{Fig02VlVb}. 

\section{Method} \label{sec:method}

The members of the same OB association exhibit similar positions and kinematics, although they are in unbound systems \citep{Wright2023ASPC..534..129W}. Therefore, we can identify the OB associations in the 6D phase space using LAMOST OB stars. \citet{Yang2019ApJ...880...65Y} developed a statistic that focuses on the incidence of close pairs in ($l, b, d, V_{\rm los}, V_{\rm l} ,V_b$), based on the clustering estimator from \citet{Starkenburg2009ApJ...698..567S} and \citet{Janesh2016ApJ...816...80J}. Following the method of \citet{Yang2019ApJ...880...65Y}, we define the separation between two stars in the six-dimensional space ($l, b, d, V_{\rm los}, pmra ,pmdec$). We then apply the friends-of-friends (FoF) algorithm to identify groups of stars that are likely to be physically associated.

\subsection{6D Distances}

The “6D distances" between any two stars are calculated using their 6D ($l, b, d, V_{\rm los}, pmra ,pmdec$) information. Here, $l$ and $b$ represent Galactic longitude and Galactic latitude in the Galactic coordinate system, $d$ is the distance to the Sun, $V_{\rm los}$ is the line-of-sight velocity, and (pmra, pmdec) are proper motion. $V_{\rm los}$ is measured in the GSR frame. The ``6D Distances" between any two stars, $i$ and $j$, are defined as follows: 

\begin{equation}\label{equa1}
\begin{aligned}
    \delta^{2}_{6D}=\Big\{\omega_\theta\theta^{2}_{ij}+\omega_{\Delta d}(d_i-d_j)^{2}+\omega_{\Delta V_{\rm los}}(V_{{\rm los},i}-V_{{\rm los},j})^{2}+\\ 
     \omega_{\Delta pmra}(pmra_i-pmra_j)^{2}+\omega_{\Delta pmdec}(pmdec_i-pmdec_j)^{2}\Big\} 
\end{aligned}
\end{equation}
here, $\theta_{ij}$ represents the great circle distance between two stars, and is calculated as
\begin{align}
    {\rm cos}\theta_{ij}=({\rm cos}b_i {\rm cos}b_j {\rm cos} (l_i-l_j)+{\rm sin}b_i {\rm sin}b_j)(\frac{d_j+d_j}{2})
\end{align}
$\omega_\theta$, $\omega_{\Delta d}$, $\omega_{\Delta V_{\rm los}}$, $\omega_{\Delta pmra}$, and $\omega_{\Delta pmdec}$ are weights used to normalize the corresponding components. They are defined as follows:

\begin{equation}\label{eqution3}
\begin{aligned}
    \omega_\theta=\frac{1}{(\theta_{90})^2} \\
    \omega_{\Delta d}=\frac{1}{(\Delta d_{90})^2}\frac{({(\frac{d_{\rm err}(i)}{d(i)})^2}+({\frac{d_{\rm err}(j)}{d(j)})}^2)}{2{\langle \frac{d_{\rm err}}{d} \rangle}^2} \\
    \omega_{\Delta V_{\rm los}}= \frac{1}{(\Delta V_{\rm los}^{90})^2}\frac{V^{2}_{\rm los,err}(i)+V^{2}_{\rm los,err}(j)}{2\langle V_{\rm los,err} \rangle}\\
    \omega_{\Delta pmra}= \frac{1}{(\Delta pmra^{90})^2}\frac{(pmra_{\rm err}(i))^2+(pmra_{\rm err}(j))^2}{2\langle pmra_{\rm err} \rangle}\\
    \omega_{\Delta pmdec}= \frac{1}{(\Delta pmdec^{90})^2}\frac{(pmdec_{\rm err}(i))^2+(pmdec_{\rm err}(j))^2}{2\langle pmdec_{\rm err} \rangle}
\end{aligned}
\end{equation}

In equation \ref{eqution3}, $\theta_{90}$, $\Delta d_{90}$, $\Delta V_{\rm los}^{90}$, $\Delta pmra^{90}$, and $\Delta pmdec^{90}$ represent the values at the 90th percentile of the angular separation distances ($\theta$), the heliocentric distance separation ($\Delta d$), line-of-sight velocity separation ($\Delta V_{\rm los}$), pmra separation ($\Delta pmra$), and pmdec separation ($\Delta pmdec$) distribution, respectively. $\langle \frac{d_{\rm err}}{d} \rangle$, $\langle V_{\rm los,err} \rangle$, $\langle pmra_{\rm err} \rangle$, and $\langle pmdec_{\rm err} \rangle$ refers to the median value of all stars. To normalize the corresponding components, we use the values at the 90th percentile of $\theta$, $\Delta d$, $\Delta V_{\rm los}$,$\Delta pmra$, and $\Delta pmdec$ distributions, and the median values of  $\frac{d_{\rm err}}{d}$, $V_{\rm los,err}$, $ pmra_{\rm err}$, and $pmdec_{\rm err}$, to avoid abnormal values. Finally, we calculate the ``6D distances" for 19,933 OB stars using the method described above.

\subsection{Clustering Algorithm: FoF}

\begin{figure}[htp!]
    \centering
    \includegraphics[width=0.45\textwidth]{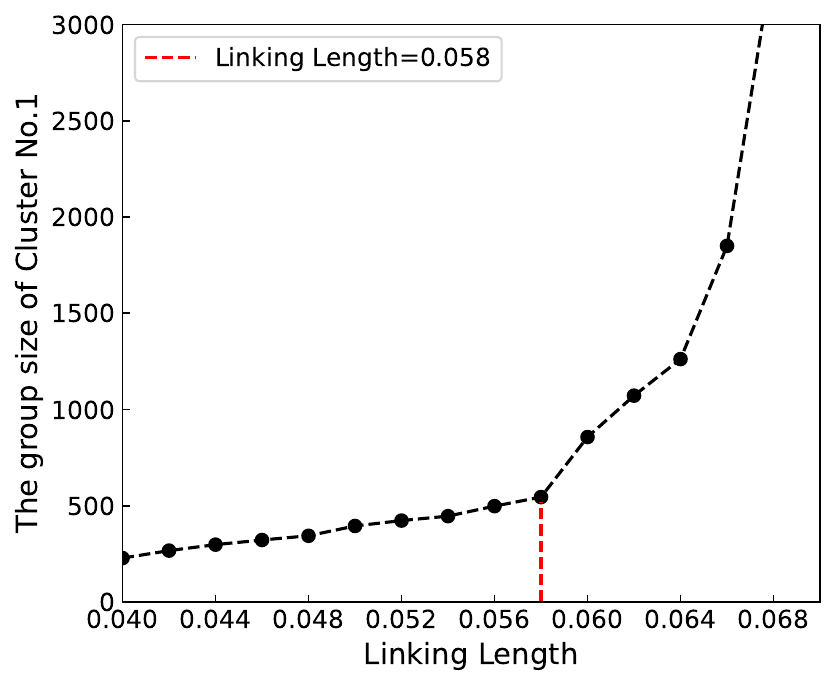}
    \caption{Distribution of group size of Cluster No.1 at different linking lengths. Here, the group size represents the number of OB stars in the cluster. Cluster No.1 is the group with the largest number of member stars.}
    \label{fig03:Groupssize}
\end{figure}

\begin{figure*}[htp!]
    \centering
    \includegraphics[width=0.9\textwidth]{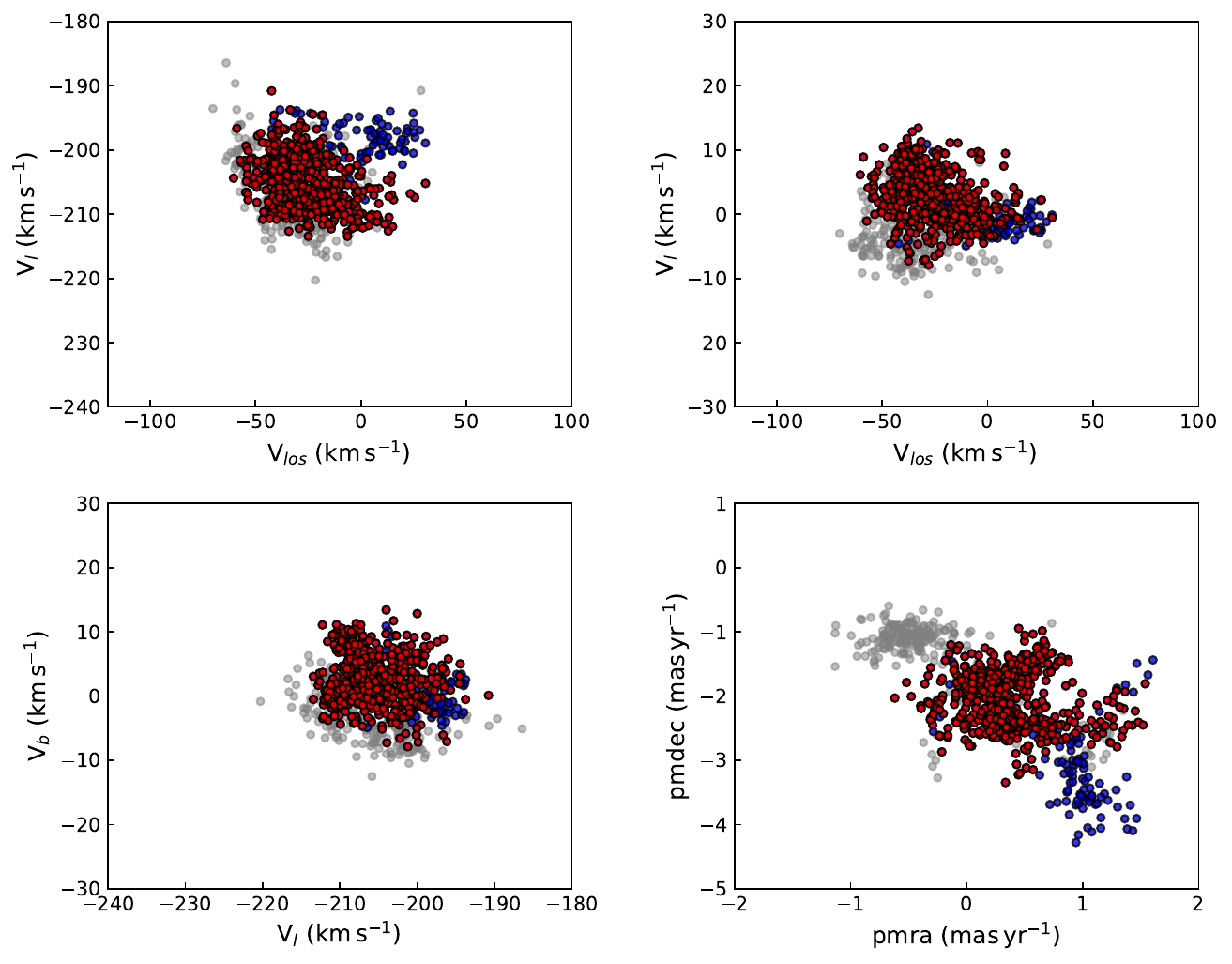}
    \caption{Distribution in clustering parameter phase space of the Cluster No.1 members at linking length 0.058 (red), 0.059 (blue), and 0.060 (gray). }
    \label{fig04:pmrapanel}
\end{figure*}

To assign OB stars with similar characteristics into groups, we employ the FoF algorithm. In this algorithm, the groups are built by containing all stars with “6D distances" less than a certain threshold, called the linking length \citep{Yang2019ApJ...880...65Y}. As shown in Figure \ref{Fig02VlVb}, most OB stars in our sample have Galactocentric distances ranging from 7 to 14\,kpc. When we set a small linking length, the FoF algorithm can only identify a few groups. Conversely, when we set a large linking length, the FoF algorithm can also find a few groups due to the merged groups. To determine a suitable linking length to assign our OB stars into groups, we study the changing trend of the group size of the maximum group when the linking length increases. 

Figure \ref{fig03:Groupssize} shows the group size \footnote{Here, the group size represents the numbers of OB stars in the cluster} of the largest group No.1 changes as the linking length increases. The group size initially grows slowly and then jumps up at a certain linking length of 0.058. Since this rapid increase could correspond to a merger of two or more groups, we checked if the group showed a multimodal distribution after the jump. Figure \ref{fig04:pmrapanel} shows the distribution of the largest group at different linking lengths (0.058, 0.059, and 0.060) in the velocity and proper motion plane. The results suggest that the group merges multiple components when applying a linking length of 0.059, especially in the proper motion space.

Therefore, we adopted a linking length of 0.058 before the jump and identified 179 groups containing at least 5 OB stars (min$\_$group$\_$size = 5). In the FoF algorithm, the min$\_$group$\_$size, which represents the minimum number of stars required for a grouping to be considered as a cluster, is a relatively intuitive parameter selection. To minimize the contamination from field stars as association members, we selected groups with min$\_$group$\_$size=10 as OB associations and groups with 5\,$\leq$\,group size\,$\leq$\,9 as OB association candidates \citep{Ward2018MNRAS.475.5659W,Chemel2022MNRAS.515.4359C}. Finally, we identify 67 OB associations and 112 OB association candidates and list their parameters and the parameters of their members in Table \ref{Table1}. To test the reliability of the FoF method for identifying OB associations, we have completed a test in Appendix \ref{AppendC}. The average coverage and purity of 14 simulated OB associations with group size $\geq$ 50 and 53 simulated OB associations with 10 $\leq$ group size $<$ 50 identified by the FoF method are 97.7\%, 87\%, 71.3\%, and 75.3\%, respectively. The Bhattacharyya coefficient calculated using the 3D positions ($X$, $Y$, $Z$) and 3D velocities ($V_{\rm x}$, $V_{\rm y}$, $V_{\rm z}$) of Cluster No.1 and simulated Cluster No.1 indicates that the simulated association constructed using 0.8$\sigma$ is the closest to our real OB associations. The coverage and purity of simulated OB associations with 0.8$\sigma$ identified using the FoF method are 100\% and 87\% for 14 simulated OB associations with group size $\geq$ 50, 67\% and 76\% for 53 simulated OB associations, respectively. The simulated results indicate that the FoF method is an effective way to identify OB associations.

\subsection{Galactic rotation curve}

The Galactic rotation curve is practically flat by analyzing the line-of-sight velocities and proper motions of OB associations \citep{Mel2009MNRAS.400..518M,mel2017MNRAS.472.3887M}. The angular rotation velocity at the solar distance has different values using different tracers. A large angular rotation velocity of $\Omega_0$=31$\pm$1\,km\,s$^{-1}$\,kpc$^{-1}$ at the solar distance is obtained by studying OB associations with the \textit{Hipparcos} data \citep{Mel2009MNRAS.400..518M}. \citet{Zabolotskikh2002AstL...28..454Z} derived a value of $\Omega_0$=29.6$\pm$1.6\,km\,s$^{-1}$\,kpc$^{-1}$, based on the kinematics of blue supergiants. Similarly, \citet{Bobylev2018AstL...44..676B} and \citet{Bobylev2022AstL...48..243B} used OB stars to obtain angular rotation velocities of $\Omega_0$=28.92$\pm$0.39\,km\,s$^{-1}$\,kpc$^{-1}$ and $\Omega_0$=29.20$\pm$0.18\,km\,s$^{-1}$\,kpc$^{-1}$, respectively.

We use the method described by \citet{Bobylev2022AstL...48..243B} to determine the parameters of the Galactic rotation curve. Briefly, based on the equations derived from Bottlinger's formulas, in which the angular velocity $\Omega$ is expanded into a series up to the second-order terms in r/R$_0$:

\begin{multline}
    V_r=-U_\odot {\rm cos}b{\rm cos}l-V_\odot {\rm cos}b {\rm sin}l -W_\odot {\rm sin}b+R_0(R-R_0) \\ {\rm sin}l{\rm cos}b\Omega_0^{'}+\frac{1}{2}R_0(R-R_0)^{2}{\rm sin}l{\rm cos}b\Omega_0^{''}
\end{multline}
\begin{multline}
    V_l=U_\odot {\rm sin}l-V_\odot {\rm cos}l -r\Omega_0{\rm cos}b+(R-R_0)(R_0{\rm cos}l-r{\rm cos}b) \\\Omega_0^{'}+\frac{1}{2}(R-R_0)^{2}(R_0{\rm cos}l-r{\rm cos}b)\Omega_0^{''}
\end{multline}
\begin{multline}
    V_b=U_\odot {\rm cos}b{\rm sin}l-V_\odot {\rm sin}l{\rm sin}b -W_\odot {\rm cos}b-R_0(R-R_0) \\ {\rm sin}l{\rm sin}b\Omega_0^{'}-\frac{1}{2}R_0(R-R_0)^{2}{\rm sin}l{\rm sin}b\Omega_0^{''}
\end{multline}

\begin{equation}
 R^2=r^2{\rm cos}^2b-2R_0r{\rm cos}b{\rm cos}l+R_0^2   
\end{equation}

\begin{table*}[htp!]
\centering
\caption{The parameters of 67 OB associations and 112 OB association candidates. The parameters from left to right are OB associations number, the average Galactic coordinates ($l$ and $b$), the average proper motion (pmra and pmdec), the average velocity ($V_{\rm los}$, $V_l$, and $V_b$), the velocity dispersion ($\sigma_{V_{\rm los}}$, $\sigma_{V_l}$, and $\sigma_{V_b}$), and average heliocentric distance and uncertainties. The total number of OB stars (Group size) in the OB associations or OB association candidates and sizes are also shown in this table. The velocity dispersion  and heliocentric distance uncertainty are calculated using the standard deviation. This full Table is available in its entirety in FITS format.}\label{Table1}
\begin{tabular}{lrrrrrrrrrrrrr}
\hline
  \multicolumn{1}{c}{Cluster} &
  \multicolumn{1}{c}{$l$} &
  \multicolumn{1}{c}{$b$} &
  \multicolumn{1}{c}{pmra} &
  \multicolumn{1}{c}{pmdec} &
  \multicolumn{1}{c}{$V_{\rm los}$} &
  \multicolumn{1}{c}{$\sigma_{V_{\rm los}}$} &
  \multicolumn{1}{c}{$V_l$} &
  \multicolumn{1}{c}{$\sigma_{V_l}$} &
  \multicolumn{1}{c}{$V_b$} &
  \multicolumn{1}{c}{$\sigma_{V_b}$} &
  \multicolumn{1}{c}{$d_{\rm Sun}$}&
  \multicolumn{1}{c}{Group size$^a$} &
  \multicolumn{1}{c}{Size$^b$} \\
  \cline{6-11}
  \multicolumn{1}{c}{} &
  \multicolumn{2}{c}{(deg)} &
  \multicolumn{2}{c}{(mas yr$^{-1}$)} &
  \multicolumn{6}{c}{(km s$^{-1}$)} &
  \multicolumn{1}{c}{(pc)}&
  \multicolumn{1}{c}{}&
  \multicolumn{1}{c}{(pc)}\\
\hline
\multicolumn{14}{c}{67 OB associations (Cluster No.1-67)}\\
\hline
  1 & 187.59 & 1.63 & 0.36 & -2.06 & -25.3 & 15.6 & -205.28 & 4.09 & 2.66 & 3.9&1749$\pm144$ & 545(6) & 158\\
  2 & 134.67 & -3.83 & -0.69 & -1.17 & 102.66 & 10.39 & -168.19 & 2.89 & 3.43 & 2.65 &2233$\pm153$& 203(0) & 106\\
  3 & 152.27 & -1.7 & 0.76 & -1.56 & 73.21 & 10.36 & -191.7 & 6.22 & 5.25 & 5.61&1540$\pm420$ & 173(0)  & 192\\
  \nodata & \nodata & \nodata & \nodata & \nodata & \nodata & \nodata & \nodata& \nodata& \nodata& \nodata& \nodata & \nodata& \nodata \\
  \nodata & \nodata & \nodata & \nodata & \nodata & \nodata & \nodata & \nodata& \nodata& \nodata& \nodata  & \nodata& \nodata& \nodata\\
  \nodata & \nodata & \nodata & \nodata & \nodata & \nodata & \nodata & \nodata  & \nodata& \nodata& \nodata& \nodata& \nodata& \nodata\\
  66 & 210.9 & -0.56 & -1.11 & 0.16 & -86.84 & 4.07 & -192.32 & 6.76 & -0.26 & 3.7&1431$\pm71$ & 10(0)  & 131\\
  67 & 212.22 & 0.2 & -1.12 & -1.13 & -102.06 & 6.41 & -182.38 & 6.42 & 0.01 & 9.76&1091$\pm129$ & 10(0)  & 137\\
  \hline
  \multicolumn{14}{c}{112 OB association candidates (Cluster No.68-179)}\\
  \hline
  68 & 182.55 & -0.12 & -0.35 & -0.98 & -17.87 & 6.83 & -221.72 & 0.56 & 3.67 & 0.45&909$\pm30$ & 9(0)& \nodata\\
  69 & 208.41 & -1.85 & -0.64 & -0.8 & -86.16 & 3.98 & -189.24 & 4.41 & -5.04 & 3.98&1928$\pm87$ & 9(0) & \nodata\\
  70 & 141.79 & -0.67 & -0.51 & -0.01 & 106.2 & 9.78 & -184.51 & 11.58 & 8.07 & 8.12&1036$\pm105$& 9(0) & \nodata\\
  \nodata & \nodata & \nodata & \nodata & \nodata & \nodata & \nodata & \nodata & \nodata& \nodata& \nodata& \nodata & \nodata& \nodata\\
  \nodata & \nodata & \nodata & \nodata & \nodata & \nodata & \nodata & \nodata  & \nodata& \nodata& \nodata& \nodata& \nodata& \nodata\\
  \nodata & \nodata & \nodata & \nodata & \nodata & \nodata & \nodata & \nodata  & \nodata& \nodata& \nodata& \nodata& \nodata& \nodata\\
  178 & 149.01 & -4.4 & -0.47 & -2.7 & 83.36 & 2.54 & -188.53 & 1.62 & -3.18 & 1.36&1614$\pm69$ & 5(0)& \nodata\\
  179 & 77.52 & 2.22 & -2.17 & -7.69 & 214.19 & 3.78 & 1.47 & 9.84 & -13.89 & 9.21&1040$\pm 63$ & 5(0) & \nodata\\
\hline
\multicolumn{14}{l}{a: Group size represents the total numbers of OB stars in the OB associations or OB association candidates. The numbers in } \\
\multicolumn{14}{l}{parentheses represent the number of O-type stars.}\\
\multicolumn{14}{l}{b: Size represents the diameter of OB association.}\\
\end{tabular}
\label{tab:my_label}
\end{table*}

\begin{figure}[hp!]
    \centering
    \includegraphics[width=1.0\linewidth]{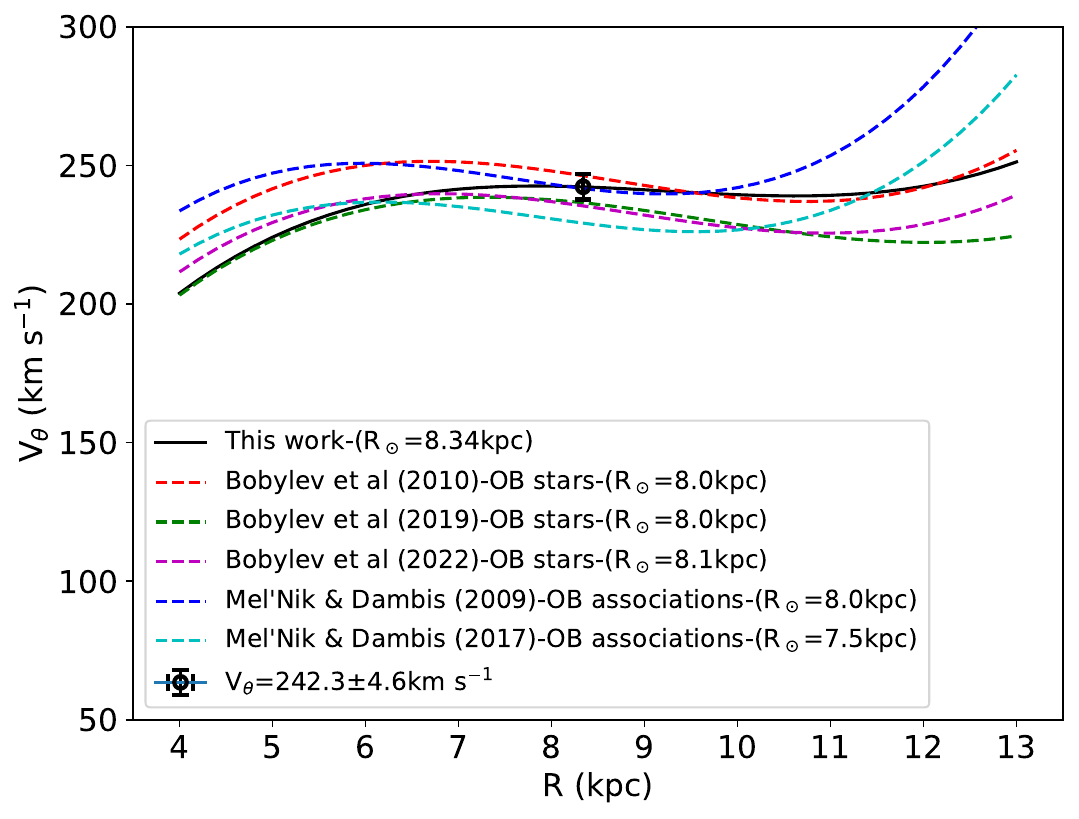}
    \caption{The Galactic rotation curves in Galactocentric distances of 4-13\,kpc, determined using OB stars and associations, assuming different $R_\odot$. Different colors represent different works, as indicated in the legend. The circular velocity (242.3$\pm$4.6\,km\,s$^{-1}$) of the Sun at 8.34\,kpc is also marked.}
    \label{fig:galactic}
\end{figure}

Where $V_r$, $V_l$, and $V_b$ represent the line-of-sight velocity, the two tangential velocity components $V_l$ = 4.74$r\mu_l$cos$b$ and $V_b$ = 4.74$r\mu_b$ along the Galactic longitude $l$ and latitude $b$, respectively. Here, $r$ is the stellar heliocentric distance in kpc. $R_0$ is the position of the Sun. $R$ is the stellar Galactocentric distance. $U_\odot$, $V_\odot$, and $W_\odot$ reflect the peculiar motion of the Sun.

We obtained $\Omega_0$=29.05$\pm0.55$, $\Omega_0^{'}$=-3.62$\pm0.15$, and $\Omega_0^{''}$=0.66$\pm0.06$ by solving the conditional equations (5)-(7) using the least-squares method, based on the 3016 member stars of 67 OB associations. The Sun is located at 8.34\,kpc. Using the rotation parameters, we estimate the circular velocity of the solar neighborhood as $V_0$=$|R_0\Omega_0|$=242.3$\pm4.6$\,km\,s$^{-1}$. In addition, we select 18,897 OB stars satisfying 7$\leq$$R_g$$\leq$13 by removing 3$\sigma$ outliers in the U, V, and W space, and obtain the $\Omega_0$=28.83$\pm0.28$ and $V_0$=240.4$\pm2.3$\,km\,s$^{-1}$. Figure \ref{fig:galactic} shows the distribution of Galactic rotation curves ranging from 4-13\,kpc. These curves were obtained using the parameters from this work and literature \citep{Bobylev2010MNRAS.408.1788B,Bobylev2019AstL...45..331B,Bobylev2022AstL...48..243B,Mel2009MNRAS.400..518M,mel2017MNRAS.472.3887M}. It is seen that the Galactic rotation curve is practically flat in 7-13\,kpc.

\section{Results and discussion}\label{sec:RandD}

\subsection{The spatial distribution of OB associations}

\begin{figure*}[htp!]
    \centering
    \includegraphics[width=0.88\textwidth]{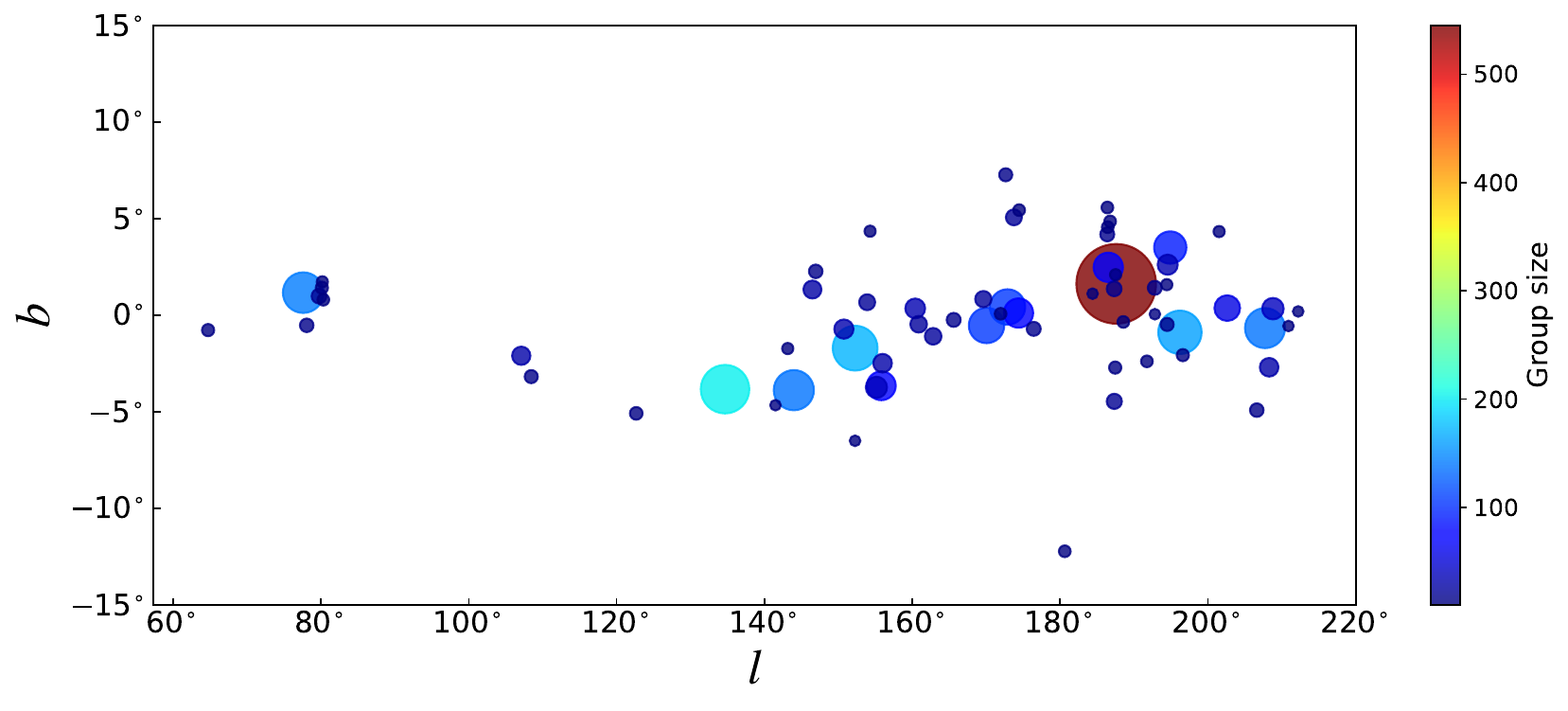}
    \caption{Distribution in Galactic coordinates ($l$ vs. $b$) of the 67 OB associations with group size larger than 9 in this work. The size and colours of the dots are proportional to the star numbers of the associations. The color bar (Group size) stands for the number of OB association members.}
    \label{fig05:GlonGlat}
\end{figure*}

\begin{figure*}[htp!]
    \centering
    \includegraphics[width=0.77\textwidth]{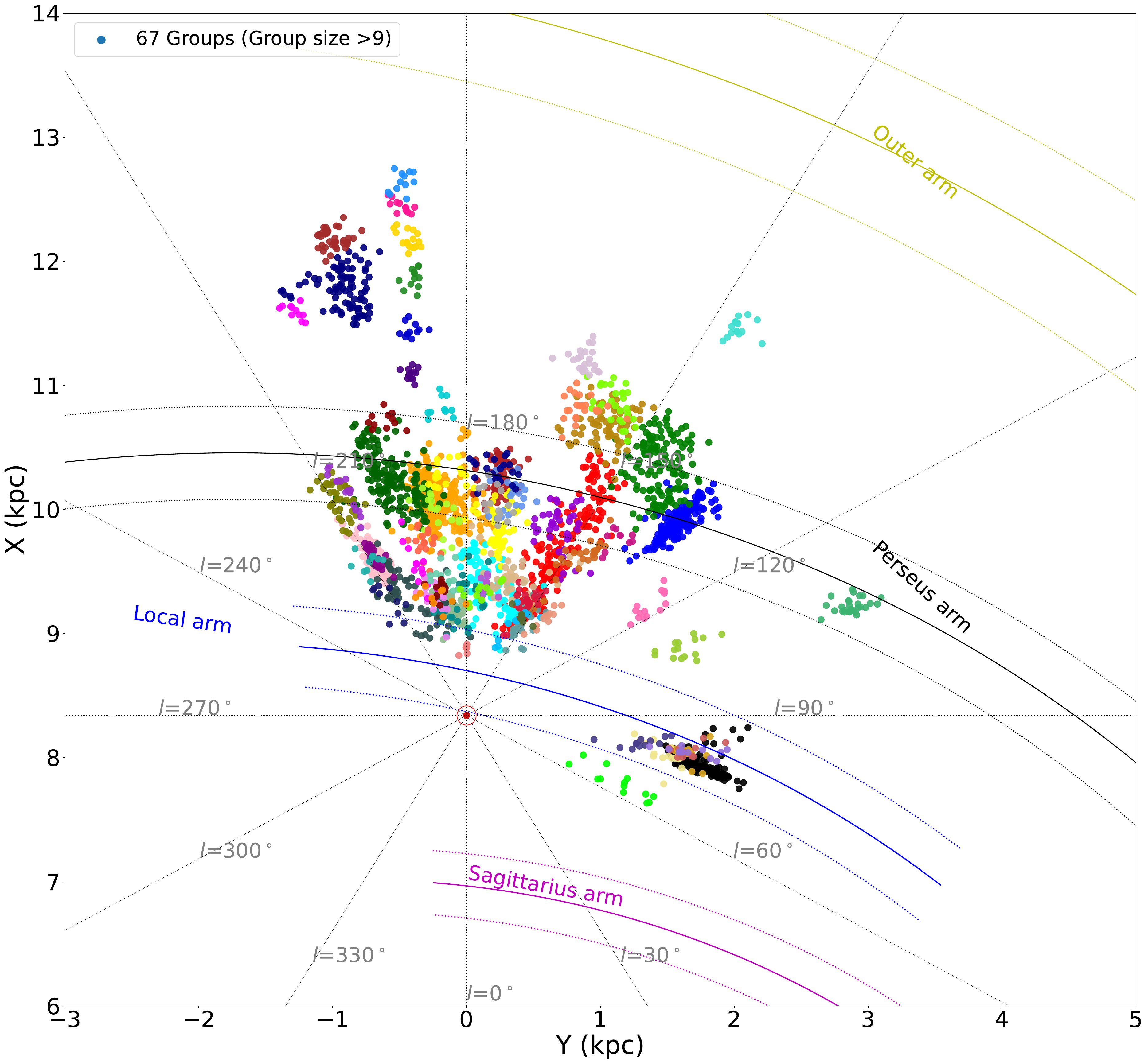}
    \caption{Distribution of the 67 OB associations with group size larger than 9 stars in the Galactic X-Y plane. The red dot circle represents the solar position (X=8.34\,kpc, Y=0\,kpc). The Galactic spiral arms are plotted using the data from \citet{Reid2014ApJ...783..130R}, while the OB associations are plotted using random colours.}
    \label{fig06:xxyy}
\end{figure*}

\begin{figure*}
    \centering
    \includegraphics[width=1.0\textwidth]{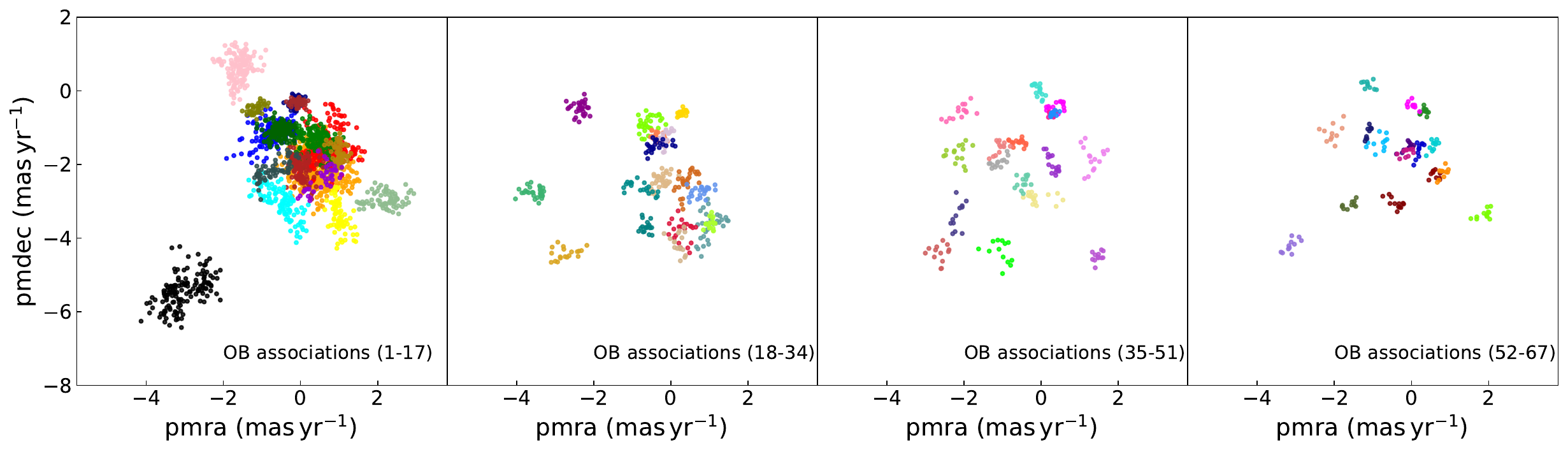}
    \caption{The panels from left to right show the distribution of the 67 OB associations in proper motion spaces. They are plotted using random colours.}
    \label{fig07:GlonGlatpmra}
\end{figure*}

\begin{figure*}
    \centering
    \includegraphics[width=0.9\textwidth]{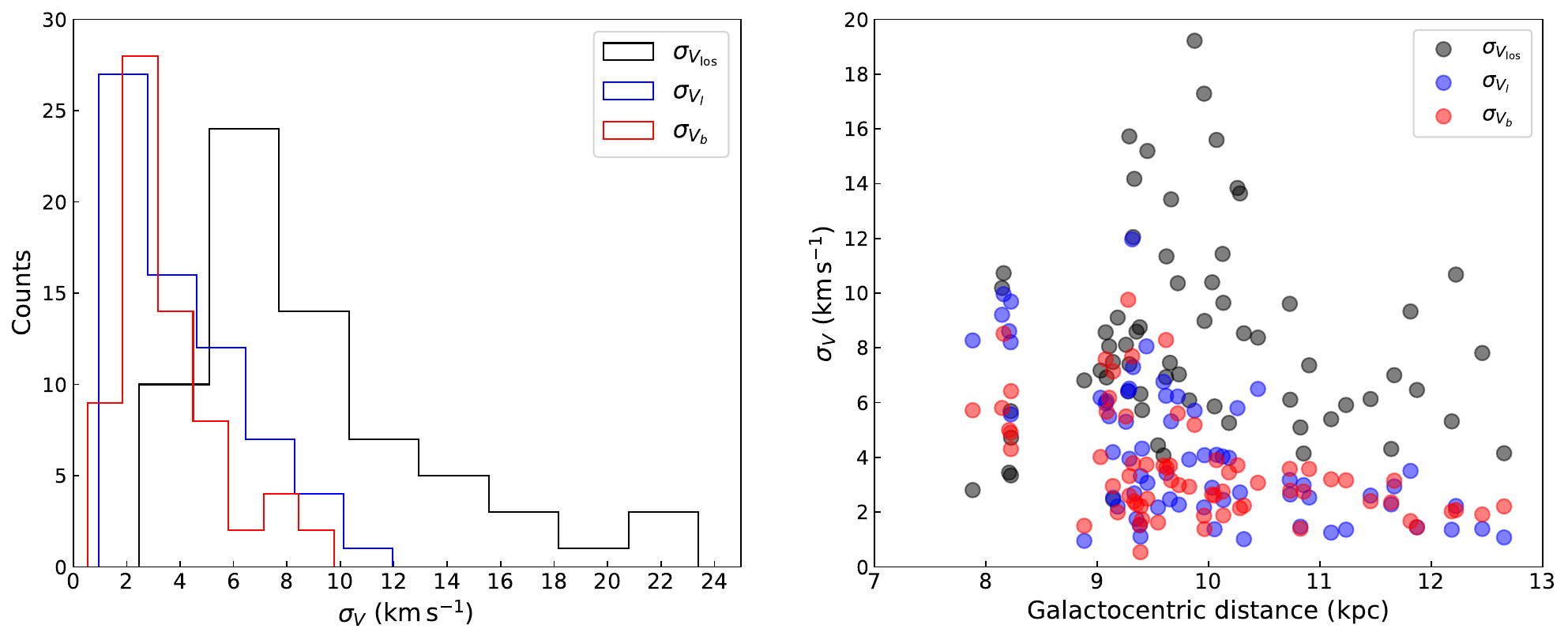}
    \caption{Left panel: the histogram shows the velocity dispersion of 67 associations in different orientations. Right panel: the velocity dispersion distribution with the Galactocentric distances of these 67 OB associations.}
    \label{fig08:GlonGlat}
\end{figure*}

In Figure \ref{fig05:GlonGlat}, we show the distribution of 67 OB associations in the Galactic latitude vs. Galactic longitude plane. The number of OB association members is represented by dots of different colors and sizes. Notably, the majority of these OB associations contain fewer than 200 members and are concentrated at low Galactic latitudes. This distribution suggests that OB stars predominantly form within the Galactic disk. Furthermore, it implies that the OB stars at high Galactic latitudes form in the Galactic disk and scatter into the Galactic halo by the binary ejection mechanism and the dynamical ejection mechanism \citep{Silva2011MNRAS.411.2596S,McEvoy2017ApJ...842...32M,Liu2023MNRAS.519..995L}.  

\begin{figure*}[htp!]
    \centering
    \includegraphics[width=1.0\textwidth]{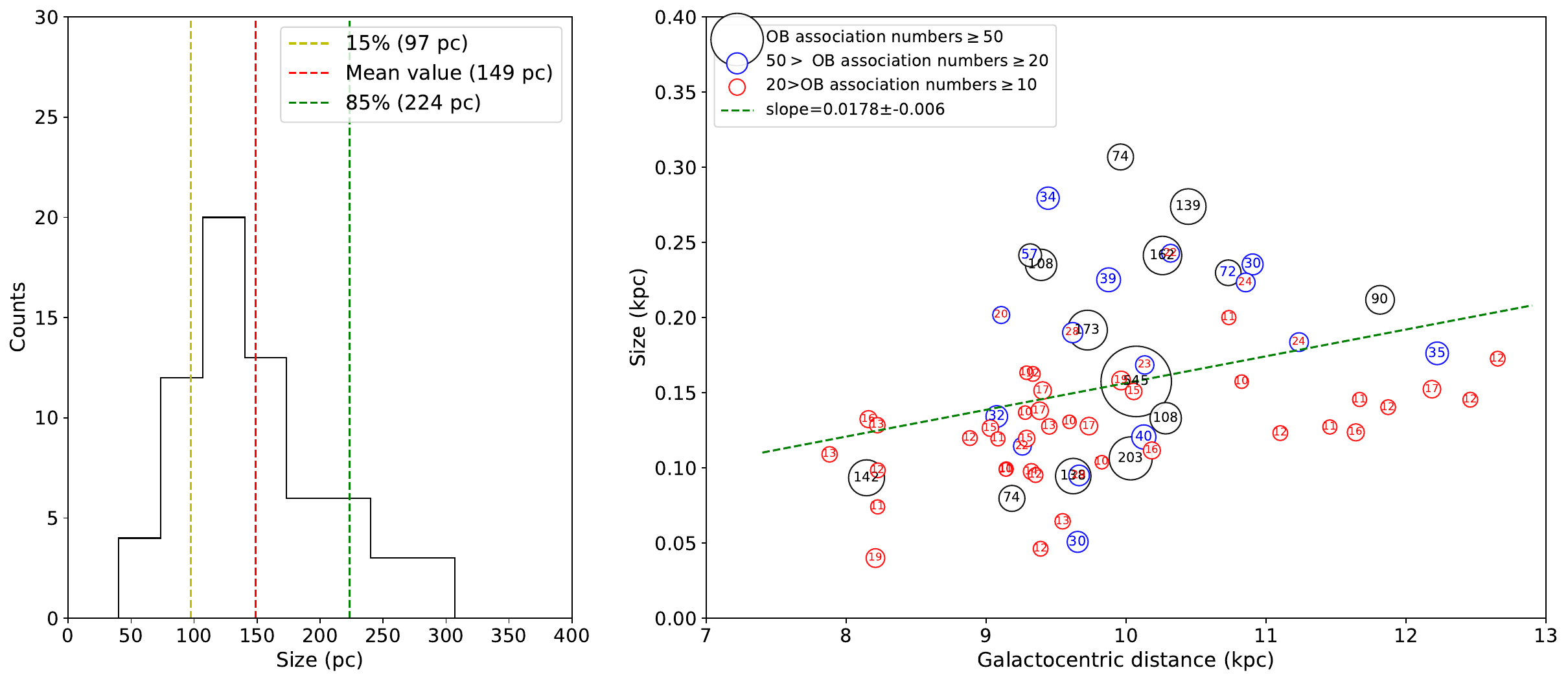}
     \caption{Left panel: the size distribution of the 67 OB associations in this work. The red dashed line represents the mean size (156\,pc).  The yellow and green dashed lines indicate the 15\% and 85\% of the size distribution (65\,pc and 249\,pc), respectively. Right panel: the size distribution of the 67 OB associations in different Galactocentric distances. The size of the unfilled circle represents the number of member stars of OB associations, and we also marked it in the figure. For example, $\textcircled{\scriptsize 30}$ represents that the number of member stars in the OB association is 30. The green dashed line represents the trend of the OB association sizes as they change with the Galactocentric distances.} 
    \label{fig00:Size}
\end{figure*}

Figure \ref{fig06:xxyy} shows the distribution of 67 OB associations in the Galactic X-Y plane, and the Galactic spiral obtained from \citet{Reid2014ApJ...783..130R} is also shown. Most OB associations with larger group sizes are located between the Perseus Arm and the Local Arm. OB associations cannot be traced to the Outer Arm and Sagittarius Arm due to the limited distance accuracy of \textit{Gaia} DR3 and selection effects from the LAMOST survey. The spatial distribution of OB associations also implies that there are star-formation regions between the Galactic spiral arms. In addition, these OB associations exhibit a larger extension in the Galactic X-direction, which can be attributed to larger distance errors in that direction. However, all OB associations show a high degree of aggregation in the pmra vs. pmdec space, as shown in Figure \ref{fig07:GlonGlatpmra}. The spatial distribution of 117 OB association candidates is also presented and discussed in Appendix \ref{AppendB}.

\subsection{The velocity dispersions of OB associations}

The velocity dispersions of OB associations represent their dynamical state and gravitational boundedness. The velocity dispersion ($\sigma_v$) is calculated using the standard deviation of velocities. The left panel of Figure \ref{fig08:GlonGlat} shows the distribution of velocity dispersion for the 67 OB associations in different directions. It can be seen that they show larger velocity dispersions along the line of sight due to their larger uncertainty on radial velocity measurements. As shown in the left panel of Figure \ref{fig08:GlonGlat}, the OB associations in this work have radial velocity dispersions ($\sigma_{V_{\rm los}}$) of 2.5$-$23.4\,km s$^{-1}$ with a median of 7.4\,km s$^{-1}$ and a mean of 8.9\,km s$^{-1}$. \citet{Ward2018MNRAS.475.5659W} studied 18 associations using proper motion data from \textit{Gaia} DR1 and measured velocity dispersions of 3$–$13\,km s$^{-1}$, with a median of 7\,km s$^{-1}$. \citet{Melnik2020MNRAS.493.2339M} measured an average velocity dispersion of 4.5 km s$^{-1}$ from 28 associations studied with \textit{Gaia} DR2. Compared to their results, OB associations in this work show larger radial velocity dispersions, which is probably attributed to the identification methods. In addition, we find that the four OB associations with the greatest velocity dispersions have large sizes. The right panel of Figure \ref{fig08:GlonGlat} shows the distribution of velocity dispersion for the 67 OB associations in different Galactocentric distances, their dispersions decrease with increased Galactocentric distances.

\subsection{Sizes of OB associations}

The sizes of OB associations vary from tens to hundreds of parsecs \citep{Blaauw1964ARA&A...2..213B,Gouliermis2018PASP..130g2001G,Wright2023ASPC..534..129W}. These sizes are also influenced by factors such as the methods used to define the borders, the membership of the included systems, and which systems are considered within the sample. In this study, we calculate the sizes (also called diameter) of the OB associations by projecting their members onto a circle of the celestial sphere, finding the smallest circle containing 68\% of the members, and then multiplying the diameter of the circle by the average heliocentric distance of OB association members \citep{Chemel2022MNRAS.515.4359C,Wright2023ASPC..534..129W}.

The left panel of Figure \ref{fig00:Size} shows the size distribution of 67 OB associations. Most OB associations have sizes ranging from 97\,pc to 224\,pc, with a mean size of about 149\,pc. \citet{Schmidt1958AN....284...76S} measured a mean size of 148\,pc, while \citet{Garmany1992A&AS...94..211G} measured the sizes of 18 Galactical OB associations, obtaining a mean size of 137\,pc. The average diameter of the OB associations measured by \citet{MelNik1995AstL...21...10M} is about 40\,pc. Compared to previous results, our OB association has a larger mean size. This increase in size can be attributed to two main factors: (i) differences in the identification methods and defined boundaries of OB associations. (ii) differences in the members of OB associations. The sizes of OB associations tend to increase with age \citep{Blaauw1964ARA&A...2..213B, Ward2018MNRAS.475.5659W}. Many OB associations also contain open or embedded clusters, such as $\gamma$ Vel in Vela OB2 \citep{Jeffries2014A&A...563A..94J}. The majority of the OB stars used to identify OB associations in this study are late B-type stars \citep{Liu2024ApJS..275...24L}. And the ages of our OB associations may be older than those from the literature. Therefore, our OB associations should naturally have a larger size and are more loosely bound than those in previous studies \citep{Chemel2022MNRAS.515.4359C}. 

The right panel of Figure \ref{fig00:Size} shows the distribution of sizes of 67 OB associations at different Galactocentric distances. It is seen that the sizes of OB associations increase with increased Galactocentric distances. This is because Galactic gas density and gravitational interaction will weaken when Galactocentric distances increase \citep{Dobbs2013MNRAS.432..653D}. As a result, the OB associations formed will be looser and have a larger size. Furthermore, we also study the significance of Galactocentric and heliocentric distances in determining the association sizes by fitting the coefficients (a, b, and c) of the following formula: log Size=a+b*log (Heliocentric distance)+c*log (Galactocentric distance). We obtain b=-0.185$\pm$0.155 and c=2.151$\pm$0.682, indicating that the Galactocentric center distance has a much more significant impact on association sizes.

\begin{figure}[htp!]
    \centering
    \includegraphics[width=0.5\textwidth]{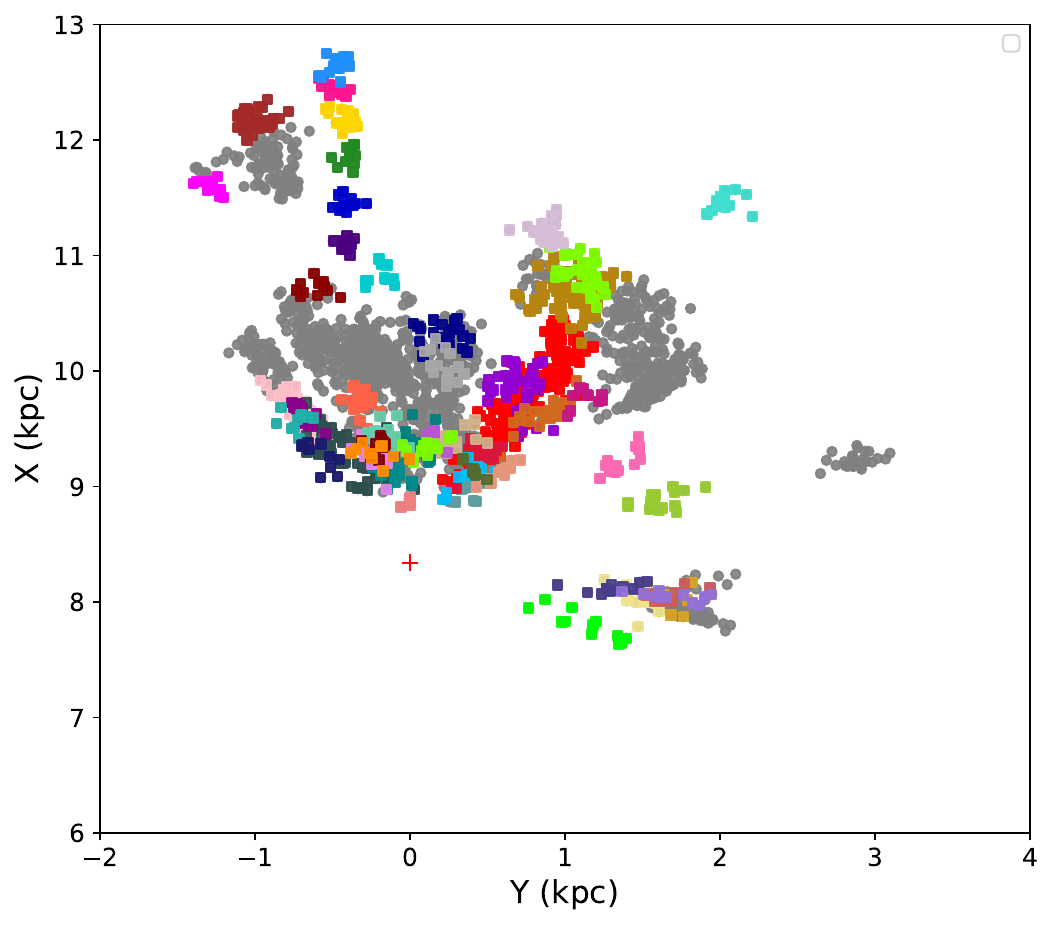}
    \caption{The distribution of 49 newly identified OB associations and 18 known OB associations in the Galactic X-Y plane. The gray dots represent the 18 common OB associations, while the differently colored squares represent the 49 new OB associations. The red cross marks the solar position (X=8.34\,kpc, Y=0\,kpc) obtained from \citet{Reid2014ApJ...783..130R}.}
    \label{fig08:Glonnew}
\end{figure}

\subsection{Comparison with \citet{Chemel2022MNRAS.515.4359C}}

\citet{Chemel2022MNRAS.515.4359C} applied the HDBSAC$^*$ clustering algorithm \citep{McInnes2017JOSS....2..205M} in the five-dimensional space (X, Y, Z, pmra, and pmdec) of the Cartesian heliocentric coordinates to search for OB associations and identified 214 clusters. We cross-matched the 67 OB association members with the 214 clusters catalog from \citet{Chemel2022MNRAS.515.4359C}, and obtained 350 common OB stars. Considering the different criteria defined for OB associations and OB association candidates in this study, we use the following criteria to define as same OB associations. For our OB associations, if there are more than three common OB stars between our sample and those from \citet{Chemel2022MNRAS.515.4359C}, they are considered to be the same. We find that 49 of 67 OB associations are newly identified.

In Figure \ref{fig08:Glonnew}, we show the distribution of the 49 newly identified OB associations and 18 OB associations identified by \citet{Chemel2022MNRAS.515.4359C} in the Galactic X-Y plane. It can be seen that the newly identified 49 OB associations have a larger spatial distribution and smaller group size compared to the previously identified associations. We also cross-matched our 112 OB association candidate members with the 214 clusters catalog from \citet{Chemel2022MNRAS.515.4359C}, and found only 49 common OB stars. For our OB association candidates, if there are more than two common OB stars between our sample and those from \citet{Chemel2022MNRAS.515.4359C}, they are considered to be the same. We find that 107 of the 112 OB association candidates are newly identified.

Compared to previous results, the sample of OB associations and OB association candidates identified using 6D OB stars in this study has more reliable member stars, especially for OB associations. The origins and dynamics of OB associations remain poorly understood, with only a small number of systems analyzed in detail. Therefore, a detailed analysis of OB associations with multidimensional parameter information will be important in revealing their formation. Moreover, by combining the improved precision of distances from parallaxes in future \textit{Gaia} data releases and low-mass young stars data identified from the LAMOST survey, we can construct the OB association samples including the young stars with different masses and multidimensional parameter information.

\section{Summary} \label{sec:conclusion}

In this work, we determine the six-dimensional (6D) information on positions, radial velocities, and proper motions for 19,933 OB stars using the \textit{Gaia} DR3 data, radial velocities derived through the cross-correlation method based on the \textit{laspec} algorithm, and distances from \citet{Bailer2021AJ....161..147B}. Based on the 6D information of OB stars and the friends-of-friends algorithm, we identified 67 OB associations and 112 OB association candidates.

The spatial distribution of the 67 OB associations indicates that they are mainly located at low Galactic latitudes, with most OB associations situated near the spiral arms of the Milky Way. While these OB associations exhibit a larger dispersion in the Galactic X-Y plane, their proper motion shows a high degree of concentration. The Galactic rotation curve, derived from the member stars of the 67 OB associations, is flat in Galactic distances of 7 to 13\,kpc. 

The distribution of velocity dispersion for most of the OB associations in this study is consistent with that of other OB associations reported in the literature. The 67 OB associations exhibit a larger size distribution than other OB associations in the Milky Way, suggesting that the OB associations in this study may have older ages \citep{Blaauw1964ARA&A...2..213B}. Our results indicate that the sizes of OB associations will increase when the Galactocentric distances increase, while their velocity dispersions decrease with increased Galactocentric distances. The 112 OB association candidates show a similar spatial distribution to the 67 OB associations. Compared to the OB associations from \citet{Chemel2022MNRAS.515.4359C}, 49 of the 67 OB associations and 107 of the 112 OB association candidates are newly identified.

In our next work, we will determine masses and ages of these OB associations by analyzing the stellar parameter analysis of their member stars. the large sample of OB associations with multidimensional parameter information will play an important role in exploring their origins and initial mass function.


\begin{acknowledgments}

 We thank the anonymous referee for the patient guidance and helpful suggestions to improve this manuscript. We also thank Dr. Chengqun Yang, Ruizhi Zhang and Jiaming Liu for technical support. This study is supported by the National Key Basic R$\&$D Program of China No. 2024YFA1611903; the National Natural Science Foundation of China under grants Nos. 11988101, 12173013, 12003045, and 12403034; the project of Hebei provincial department of science and technology under the grant No. 226Z7604G, Natural Science Foundation of Hebei Province A2024205031, Hebei Province Yan-zhao Golden Peak Talent Program (Postdoctoral Platform) for Key Talents under grant No. B2025003010, and Science Foundation of Hebei Normal University(Nos. L2024B54, L2024B55, and L2024B56). 

The Guoshoujing Telescope (the Large Sky Area Multi-Object Fiber Spectroscopic Telescope LAMOST) is a National Major Scientific Project built by the Chinese Academy of Sciences. LAMOST is operated and managed by the National Astronomical Observatories, Chinese Academy of Sciences.

This work has made use of data from the European Space Agency (ESA)  mission \textit{Gaia} (\url{https://www.cosmos.esa.int/gaia}), processed by the \textit{Gaia} Data Processing and Analysis Consortium (DPAC, \url{https://www.cosmos.esa.int/web/gaia/dpac/consortium}). Funding for the DPAC has been provided by national institutions, in particular, the institutions participating in the \textit{Gaia} Multilateral Agreement. This research has made use of the SIMBAD database, operated at CDS, Strasbourg, France.

\end{acknowledgments}

%

\vspace{5mm}
\facilities{topcat \citep{Taylor2005ASPC..347...29T}, laspec \citep{Zhang2020ApJS..246....9Z,Zhang2021ApJS..256...14Z}}





\appendix


\section{The description of Table 3}\label{AppendA}

The basic information for 3707 OB members from 67 OB associations and 112 OB association candidates is provided in the electronic version of this paper. Clusters no.1-67 and 68-179 correspond to the 67 OB associations and 112 OB association candidates, respectively.   

\begin{table*}
\normalsize
\centering
\caption{The parameters description of 3707 OB members for 67 OB associations (Cluster No.1-67) and 112 OB association candidates (Cluster No.68-179).}\label{Table2}
\begin{tabular}{llll}
\hline
\hline
\multicolumn{1}{l}{Column } &
\multicolumn{1}{l}{Format} &
\multicolumn{1}{l}{Unit} &
\multicolumn{1}{l}{Description} \\
\hline
Clusters&Double&&The name for OB associations\\
Obsid &Integer &&object's obsid from LAMOST DR7\\
Destigation&String&&object's Destigation from LAMOST DR7\\
RA&Double&deg&object's R.A. in LAMOST DR7 (J2000)\\
Dec&Double&deg&object's Dec. in LAMOST DR7 (J2000)\\
RV&Double&km s$^{-1}$&radial velocity obtained by the $laspec$ algorithm\\
RV\_err&Double&km s$^{-1}$&radial velocity error obtained by the $laspec$ algorithm\\
Source ID&Long&& \textit{Gaia} DR3 Source ID for stars\\
pmra&Double&mas yr$^{-1}$& Proper motion in right ascension direction from \textit{Gaia} DR3 \\
pmdec&Double&mas yr$^{-1}$&Proper motion in declination direction from \textit{Gaia} DR3\\
pmra\_err&Double&mas yr$^{-1}$&pmra error from \textit{Gaia} DR3\\
pmdec\_err&Double&mas yr$^{-1}$&pmdec error from \textit{Gaia} DR3\\
Glon&Double&deg&Galactic latitude\\
Glat&Double&deg&Galactic longitude\\
$V_{\rm los}$&Double&km s$^{-1}$&line-of-sight velocity in the Galactic standard of rest\\
$V_l$&Double&km s$^{-1}$&Galactic longitude velocity in the Galactic standard of rest\\
$V_b$&Double&km s$^{-1}$&Galactic latitude velocity in the Galactic standard of rest\\
$X$&Double&kpc&Galactocentric coordinate points to the direction opposite to that of the Sun\\
$Y$&Double&kpc&Galactocentric coordinate points to the direction of Galactic rotation\\
$Z$&Double&kpc&Galactocentric coordinate points toward the north Galactic pole\\
rpgeo&Double&pc& Median of the photogeometric distance posterior from \citet{Bailer2021AJ....161..147B}\\ 
B\_rpgeo&Double&pc& 84th percentile of the photogeometric distance posterior\\ 
b\_rpgeo&Double&pc& 16th percentile of the photogeometric distance posterior \\ 
Spectral type&Character&& The spectral types of O-type and B-type stars are obtained from \citet{Liu2024ApJS..275...24L} \\
Comment &String& &These stars can be found in LAMOST DR3, DR5 or DR7. \\
\hline
\multicolumn{4}{l}{(This table is available in its entirety in FITS format.)}\\
\end{tabular}
\end{table*}
\section{The spatial distribution of 112 OB association candidates}\label{AppendB}

\begin{figure*}
    \centering
    \includegraphics[width=1.0\textwidth]{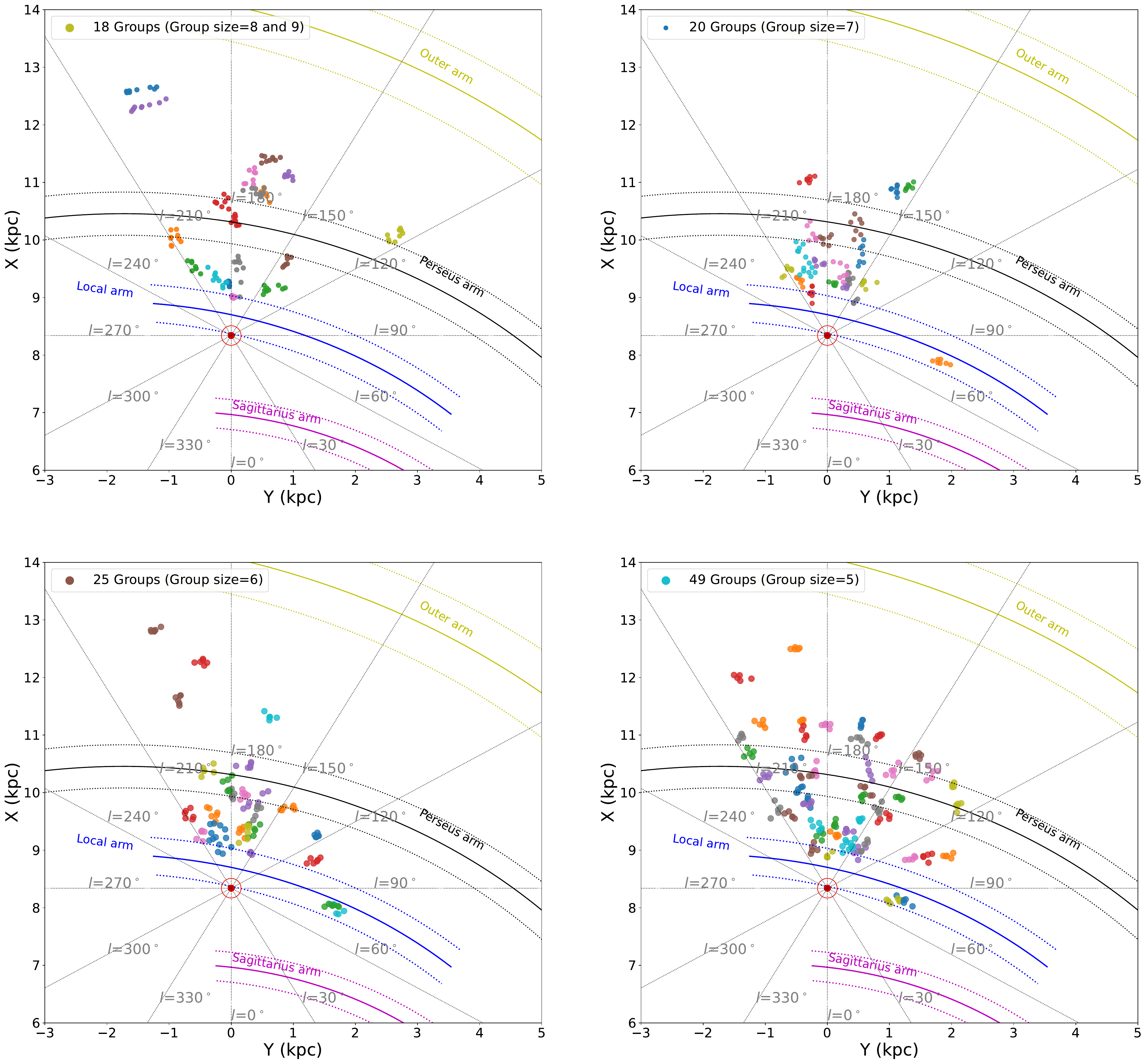}
    \caption{Similar to Figure \ref{fig06:xxyy}, the distribution of 112 OB association candidates with group size greater than 5 and less than 10 stars in Galactic X-Y plane.}
    \label{fig09:XYpanel}
\end{figure*}

We divided the 112 OB association candidates into four different groups of different group sizes, as shown in Figure \ref{fig09:XYpanel}. It is seen that they show a similar spatial distribution to that of the 67 OB associations, and the numbers and Galactocentric distances of OB association candidates will increase with decreased group size due to the FoF clustering algorithm. Most OB association candidates are located near the spiral arms of the Milky Way. 

\section{Reliability Test of FoF method}\label{AppendC}

To validate the reliability of the FoF method for identifying OB associations and reveal various biases and artifacts, we randomly generated the 67 simulated OB associations based on the distribution of the true data of our 67 OB associations. The 67 simulated OB associations include 14 OB associations with group size $\geq$ 50 and 53 OB associations with 10 $\leq$ group size $<$ 50. The 6D parameters ($l$, $b$, $d$, $V_{\rm los}$, $pmra$, $pmdec$) of each simulated OB association members are Gaussian scattered based on the average and standard deviation ($\sigma$) of the 6D parameters of the real OB association, and the number of each simulated association members is equal to the number of the real association members.

The simulated OB associations will have a lower density than real OB associations because simulated OB association members have a larger range of parameters using the average and $\sigma$ of the 6D parameters of the real OB association. Therefore, we generate the simulated OB association members using the different values ranging from 0.5\,$\sigma$ to $\sigma$. The simulation data consists of the 67 simulated OB association members and the remainder 18,791 real OB stars. We use the FoF method to identify OB associations and repeat the above procedure 10 times for different $\sigma$ values.

\begin{figure*}
    \centering
    \includegraphics[width=1.0\textwidth]{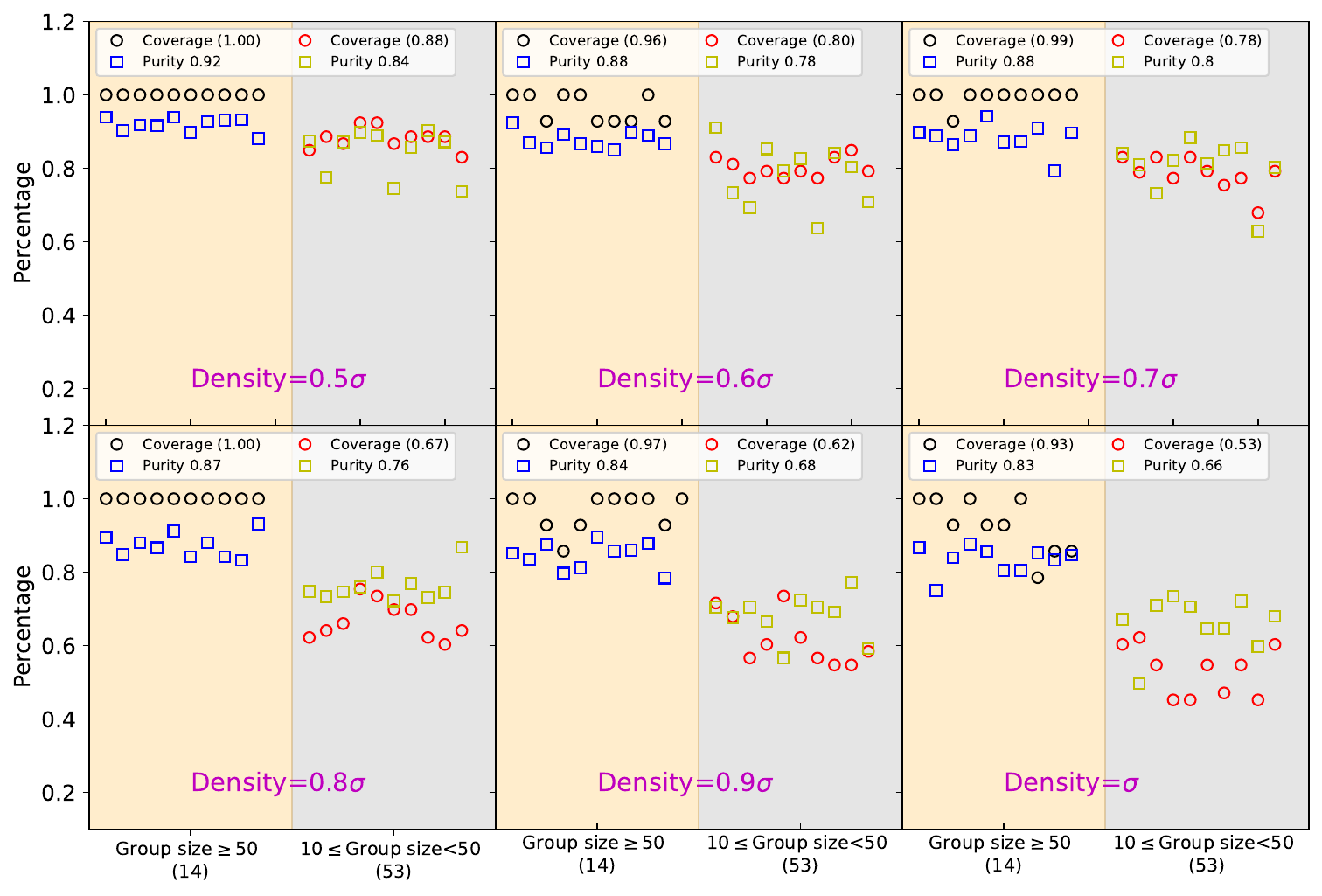}
    \caption{The distribution of coverage and purity of 14 simulated OB associations with group size $\geq$ 50 and 53 simulated OB associations with 10 $\leq$ group size $<$ 50 obtained by using different $\sigma$ values. Coverage means how many of them are located in the result. Purity is the group members left of a total number of the stars in the process. The left panel of each subgraph shows the distribution of coverage and purity of 14 simulated OB associations with group size $\geq$ 50, , while the right panel corresponds to the 53 simulated OB associations with 10 $\leq$ group size $<$ 50. Mean values of coverage and purity are also marked in the figure.}
    \label{fig14moni}
\end{figure*}

Figure \ref{fig14moni} shows the distribution of coverage and purity of simulated results using different $\sigma$ values. When the $\sigma$ value changes from 0.5\,$\sigma$ to $\sigma$, the coverage of simulated OB associations and purity of the simulated OB association members gradually decrease. The mean coverage and purity of simulated OB associations identified using the FoF method are 97.7\% and 87\% for 14 simulated OB associations with group size $\geq$ 50, 71.3\% and 75.3\% for 53 simulated OB associations, respectively. The main reason affecting the purity of the 53 simulated star associations is their relatively small number of member stars. Such as, for an OB star association with 12 member stars, when we conduct a re-simulation, the true 6D distance between them will also change. If only 9 stars are found when evaluating the simulation results, then it will not be regarded as an OB star association.

\begin{figure}
    \centering
    \includegraphics[width=0.5\linewidth]{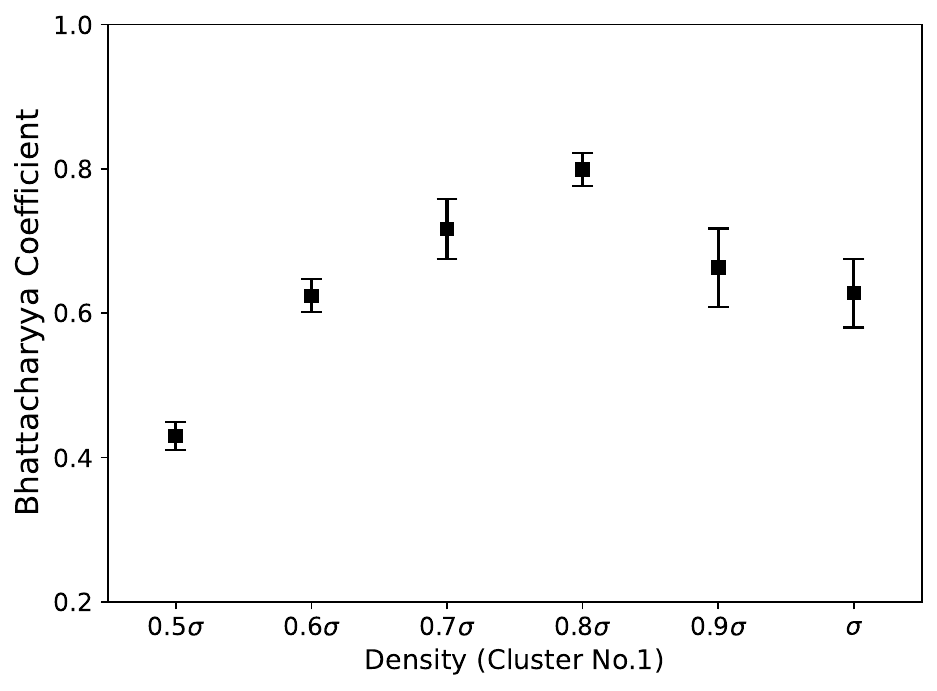}
    \caption{The distribution of Bhattacharyya coefficient for Cluster No.1. The higher the Batatahari coefficient is, the closer the distribution of the simulated associations is to the distribution of real association.}
    \label{fig15xishu}
\end{figure}

In addition, we calculate the Bhattacharyya coefficient for Cluster No.1, based on the 3D positions ($X$, $Y$, $Z$) and 3D velocities ($V_{\rm x}$,$V_{\rm y}$,$V_{\rm z}$) of true Cluster No.1 and simulated Cluster No.1. The simulated Cluster No.1 is constructed by using the different values ranging from 0.5\,$\sigma$ to $\sigma$. Figure~\ref{fig15xishu} shows the distribution of Bhattacharyya coefficient for Cluster No.1. The Bhattacharyya coefficient of Cluster No.1 is obtained using the real data of Cluster N0.1 and simulated Cluster No.1. The larger the Batychev coefficient is, the closer the distribution of the simulated association is to that of the real association. Our result indicates that OB associations constructed using the 0.8$\sigma$ are the closest to the actual distribution of real OB associations. The coverage and purity of simulated OB associations with 0.8$\sigma$ identified using the FoF method are 100\% and 87\% for 14 simulated OB associations with group size $\geq$ 50, 67\% and 76\% for 53 simulated OB associations, respectively.

\bibliography{sample631}{}
\bibliographystyle{aasjournal}



\end{document}